\documentclass{amsart}

\usepackage{t1enc,amsthm,amssymb,array,epsfig}
\newtheorem{lema}{Lemma}
\newtheorem{prop}[lema]{Proposition}
\newtheorem{teo}[lema]{Theorem}
\newtheorem{rem}[lema]{Remark}
\newtheorem{corollary}[lema]{Corollary}
\theoremstyle{definition}
\newtheorem{defin}{Definition}

\newcommand{\R}{\ensuremath{\mathbb{R}}}
\DeclareMathOperator{\Bola}{B}

\DeclareMathOperator{\Vol}{Vol}
\newcommand{\vol}[1]{\Vol(#1)}

\newcommand{\schr}{Schr\"oedinger}
\newcommand{\dis}{\displaystyle}
\newcommand{\Gd}{\mathcal{G}_d}
\newcommand{\OF}{\Omega_F}
\newcommand{\OFn}{\Omega_{F_n}}
\newcommand{\cd}{\ensuremath{c_d}}
\newcommand{\Nclass}{\ensuremath{\mathcal{N}_\gamma(\Gd;\R^n)}}
\DeclareMathOperator{\capac}{cap}

\begin{document}
\date{compiled \today}

\title [Analysis of polynomial potentials and ABJ/M-type theories]{Spectral analysis of polynomial potentials and its relation with ABJ/M-type theories}
\author [M.P. Garc\'ia del Moral,I. Martin, L. Navarro, A. J. P\'erez A. and A. Restuccia]{M.P. Garc\'ia del Moral$^1$, I.Martin $^2$, L. Navarro $^3$, A.J. P\'erez A.$^4$ and A.Restuccia$^5$}
\address{$^1$ Departamento de F\'isica, Universidad de Oviedo, Calvo Sotelo 18, 33007, Oviedo, Spain. }
\email{garciamormaria@uniovi.es}
\address{$^2$ Departamento de F\'isica, Universidad Sim\'on Bol\'ivar, Apartado 89000, Caracas 1080-A, Venezuela. }
\email{isbeliam@usb.ve}
\address{$^3$ Departamento de Matem\'aticas, Universidad Sim\'on Bol\'ivar, Apartado 89000, Caracas 1080-A, Venezuela.}
\email{lnavarro@ma.usb.ve}
\address{$^4$ Departamento de Matem\'aticas, Universidad Sim\'on Bol\'ivar, Apartado 89000, Caracas 1080-A, Venezuela.}
\email{ajperez@ma.usb.ve}
\address{$^5$ Departamento de F\'isica, Universidad Sim\'on Bol\'ivar, Apartado 89000, Caracas 1080-A, Venezuela. }
\email{arestu@usb.ve} 
\maketitle 

\begin{abstract}
We obtain a general class of polynomial potentials for which the Schr\"oedinger operator has a discrete spectrum. 
This class includes all the scalar potentials in membrane, 5-brane, p-branes, multiple M2 branes, BLG and ABJM theories. We provide a proof of the discreteness of the spectrum of the associated Schr\"oedinger operators. This a a first step in order to analyze BLG and ABJM supersymmetric theories from a non-perturbative point of view.
\end{abstract}

%%%%%%%%%%%%%%%%%%%%%%%%%%%%%%%%%%%%%%%%%% INTRODUCTION %%%%%%%%%%%%%%%%%%%%%%%%%%%%%%%%%%%%%%%%%%%%%

\vskip 1 cm

\section{Introduction}

There is  an intense activity in the spectral characterization of ABJM-type theories at perturbative level. These theories belong to a class of
superconformal Chern-Simons gauge theories in three dimensions with
$\mathcal{N}=6$ supersymmetry \cite{Aharony:2008ug}. The gauge group is $U(N)\times U(N)$ with Chern-Simon level $k$.  The case with
gauge group $U(N)\times U(M)$ with different gauge groups $N\ne M$, also called ABJ was considered in \cite{Aharony:2008gk}. ABJM
theories are special cases of the Gaiotto-Witten theories \cite{Gaiotto:2008sd} i.e. Superconformal Chern-Simons Theories with
$\mathcal{N}=4$, in which the supersymmetry is enhanced to $\mathcal{N}=6$. For the case $N=2$, the number of
supersymmetries is enhanced to $\mathcal{N}=8$ and it corresponds to the BLG theory, \cite{Bagger:2006sk},\cite{Bagger:2007jr} and
\cite{Gustavsson:2007vu}. In these papers, the fields are evaluated on a 3-algebra with positive inner metric, in terms of a unique
finite dimensional gauge group $SO(4)$ with twisted Chern-Simons terms. The ABJM theory or at least a sector of it, can also be
recovered from the 3-algebra formulation by relaxing the antisymmetry condition in its structure constants\cite{Bagger:2008se}. \newline

The interest of these ABJ-like theories is double: on one hand they
represent further evidence of the duality
$AdS_4/CFT_3$\cite{Maldacena:1997re}. This duality opens an
interesting window that allows to compute different aspects of
condensed matter in the strong coupling limit in the fields of
superconductivity, semiconductors, and so on, unreachable today by
other means. For recent reviews on this interesting topic, see
\cite{Hartnoll:2009sz},\cite{Herzog:2009xv}. On the other hand these
calculations also provide results about integrability and finiteness
properties of these superconformal Chern-Simons theories in the
strong coupling regime. There have been results in this
regard recently, see for example
\cite{Zarembo:2009au,Minahan:2009te,Marino:2009jd,Bianchi:2009ja,Bak,Rey}. 
For all these reasons, any non perturbative results related to these 
theories are of obvious interest.\newline

An important aspect of super-membranes, super 5-branes and
supersymmetric multiple-M2 branes refers to the quantum stability of the theory and the validity of the Feynman
kernel. A natural way to proceed is to formulate the theory on a compact base manifold, perform then a
regularization of the theory in terms of an orthonormal basis and
analyze properties of the spectrum of the associated Schr\"odinger
operator.
This procedure, to start with a field theory and analyze its properties 
by going to a regularized model, has been very useful in
 field theory, although relevant symmetries of the theory may be lost
  in the process. In the case of the $D=11$ supermembrane
  \cite{Bergshoeff:1987qx}, an important property of the regularization
  is that the area preserving diffeomorphism, the residual gauge symmetry
  of the supermembrane in the light cone gauge, gives rise to a $SU(N)$
  gauge symmetry of the regularized model \cite{Hoppe:1988gk,deWit:1988ig}.
  The gauge symmetry of the field theory is then 'represented' as the SU(N) gauge symmetry
   of the regularized model and it is not lost in the reduction to finite degrees of freedom.
   The quantum properties of the regularized model is then determined from the Schr\"odinger
    operator $-\triangle+V(x)+$fermionic terms, where the bosonic potential $V(x)$ has the expression
\begin{equation}
V(x)=\sum_i\Big[P_i(x)\Big]^2.\label{I}
\end{equation}
$P_i(x)$ is a homogeneous polynomial on the configuration variables
$x\in\R^n$. In the membrane theory $P_i(x)$ are of degree two.

An important aspect of $V(x)$, which determines the spectrum of the
associated Schr\"odinger operator, is the algebraic variety of zero potential
which extends to infinity on configuration spaces and the behavior
of the potential along that variety. In the case of the (bosonic)
membrane the distance between the walls of the valleys
 along the zero variety goes to zero as we approach  infinity, and this was
 interpreted in \cite{deWit:1988ct} as the main reason explaining
 the discreteness of the spectrum of the membrane hamiltonian: the wavefunction
 cannot escape to infinity. The potential in the transverse directions
 to the valleys behaves as the potential of an harmonic oscillator. The proof of the discreteness was done in \cite{Luscher:1982ma} where a bound
\begin{equation}
\langle\Psi,H\Psi\rangle\geq\langle\Psi,\lambda\Psi\rangle,
\end{equation}
in terms of a function $\lambda(x)$, with
$\lambda(x)\rightarrow\infty$ as $|x|\rightarrow\infty$ was
obtained. A proof of the discreteness of the spectrum of the
membrane, following an extension of the Barry Simon argument for a
toy model $V(x)=x^2y^2$ in two dimensions, was presented in
\cite{GarciadelMoral:2006tb}.

An important remark to be mentioned is that supermembrane theory is
an example of a field theory over a compact manifold which (at the
regularized level) has continuous spectrum \cite{deWit:1988ct}.
There are several related toy models which also have continuous
spectrum, see for example \cite{deWit:1988ct}. It is only when the
supermembrane is restricted by certain topological conditions,
non-trivial central charges,
 that the spectrum becomes discrete, with eigenvalues accumulating at 
 infinity
  \cite{GarciadelMoral:2001zb,Boulton:2001gz,Boulton:2002br,Boulton:2004yt,Boulton:2006mm}.\\
In order to analyse with more precision the supermembrane and super
5-brane potentials, and even more complicated potentials as in the
BLG and ABJM theories it is very useful to consider a necessary and
sufficient condition to have a discrete spectrum. This was achieved
by A.~M.~Molchanov \cite{Molchanov:1953am} and more recently
extended by V.~Maz'ya and M.~Schubin \cite{Maz'ya:2005v}. It makes
use of the mean value of the potential, in the sense of Molchanov , on
a star shaped cell $\mathcal{G}_d$, of diameter $d$. The spectrum
is discrete if and only if the mean value of the potential goes to
infinity when the distance from $\mathcal{G}_d$ to a fixed point on
configuration space goes to infinity in all possible ways. The
potential is assumed to be locally integrable and bounded from
below.

The mean value  in the sense of Molchanov, for the membrane theory,
was obtained in \cite{GarciadelMoral:2006tb} in terms of a strictly
positive definite inertia tensor for the membrane. As a consequence
the spectrum  of the hamiltonian of the membrane theory (a
regularized  $SU(N)$ model) is discrete. Estimates of the
eigenvalues may also be obtained by looking at the mean value at
finite distances. These estimations are also useful to characterize
the mass gap of Yang-Mills theories in the slow mode regime.  As an
example, in \cite{GarciadelMoral:2006tb} it was obtained a bound for
the $3+1 D$ $SU(3)$ hamiltonian of Yang-Mills theories in the slow
mode regime in terms of a hamiltonian whose spectrum and eigenvalues
are known, and its eigenfunctions are expressed in terms of Bessel
functions .\newline

%%%%%%%%%%
An analysis of the spectrum of the $D=11$, 5-brane in the light cone
gauge, using the Molchanov, Maz'ya and Schubin theorem was performed
in \cite{Decastro:2003jc},\cite{Martin:2007kp}. The spectrum of the
hamiltonian is also discrete. However, these results do not apply
directly to the BLG and ABJM multi M2 brane theories. A common
property of the potential in all cases is the form (\ref{I}), where
$P_i(x)$ have different expressions for each theory. The potential
is always a homogeneous polynomial on the configuration variables.
The first point to notice is that the form (\ref{I}) of the
potential does not imply discreteness of the spectrum of the
hamiltonian. One example, in  $\R^3$ is the following
\begin{eqnarray}
h&=&-\triangle+V_3\\
V_3&=&|x-y|^{\alpha}+|y-z|^{\beta}+|x-z|^{\gamma},
\end{eqnarray}
where $\alpha, \beta, \gamma$ are any real positive
numbers.$h$ has a continuous spectrum. More precisely, the essential spectrum
is non-trivial. We say that the spectrum is discrete when the bottom of the
essential
spectrum is at infinity. All quasi-eigenvectors are then eigenvectors.\\
In this paper, we obtain a general class of polynomials $P_i(x)$ in
(\ref{I}) for which the Sch\"odinger operator $-\triangle+V(x)$ has
a discrete spectrum and provide a proof of it. This class includes all potentials in
membrane, 5-brane, p-branes, multiple M2-branes, BLG and ABJM
theories. This a a first
step in order to analyse BLG and ABJM super-symmetric theories from
a non-perturbative point of view. In section 2, we present
preliminary results in particular we describe the Molchanov approach
to the analysis of the spectrum. In sections 3 and 4 we present the
new results. In section 5 we discuss the application to the BLG and
ABJM theories. Finally, in section 6 we present our conclusions.

%%%%%%%%%%%%%%%%%%%%%%%%%%%%%%%%%%%%%%%%%% 2. PRELIMIRARY RESULTS %%%%%%%%%%%%%%%%%%%%%%%%%%%%%%%%%%%%%%%%%%%%%%%

\section{Preliminary Results}

   K. Friedrichs (see \cite{Maz'ya:2005v} for further references) proved that the spectrum of
   the Schr\"{o}edinger operator $-\Delta+V$ in $L^2(\R^n)$ with a locally
   integrable potential $V$  is discrete provided $V(X)\to\infty$ as $\quad |X|\to\infty$.
    This is a sufficiency condition for the discreteness of the spectrum of
    the Schr\"{o}edinger operator but, of course it is not necessary.  In order to
    understand the quantum properties of the membrane, supermembrane, 5-brane
    and multiple brane theories and as a consequence of Yang-Mills theory it is useful to look
     for a necessary and sufficient condition that ensures the discreteness of the spectrum.
     That condition, in terms only of properties of the potential was discovered by A. M. Molchanov \cite{Molchanov:1953am}
      and more recently extended by Maz'ya and Schubin \cite{Maz'ya:2005v} and makes use of
      the mean value, in the sense of Molchanov, of the potential on a star shaped set
      when the distance from the set to an origin goes to infinity. It is naturally related to the Friedrichs condition, but by taking a mean value of the potential on a cell one obtains also a necessary condition.
    We will only state the V. Maz'ya and M. Schubin generalization of Molchanov theorem \cite{Maz'ya:2005v}.
     Let us give first some definitions involved in the formulation of the theorem.
\begin{defin}
Let $n\ge2$, $F\subset\R^n$ be compact, and $Lip_{c}(\R^n)$
the set of all real-valued functions with compact support satisfying a uniform Lipschitz condition in $\R^n$.
Then the Wiener's capacity of $F$ is defined by
\[\capac(F)=\capac_{\R^n}(F)=\inf\left\{\int_{\R^n}|\nabla u(x)|^2\,dx\Big|\ u\in Lip_{c}(\R^n),\ u|_F=1\right\}\]
\end{defin}
In physical terms the capacity of the set $F\subset\R^n$ is defined as the
electrostatic energy over $\mathbb{R}^{n}$ when the electrostatic
potential is set to $1$ on $F$.

\begin{defin}
Let $\Gd\subset\R^n$ be an open, bounded and star-shaped set of diameter $d$, let $\gamma\in(0,1)$. The \emph{negligibility class} \Nclass\
consists of all compact sets $F\subset\overline{\Gd}$ satisfying
$\capac(F)\leq\gamma\capac(\overline{\Gd})$.
\end{defin}
Balls and cubes in $\R^n$ are useful examples of such $\Gd$. In what follows we denote the ball of
diameter $d$ and center $x$ by $\Bola_{d}(x)$ and the $n-$dimensional Lebesgue measure by $\vol{\cdot}$.

\begin{teo}[Maz'ya and Shubin]\label{teo_mazya}
Let $V\in L^1_{\text{loc}}(\R^n)$, $V\geq0$.\\ Necessity: If the
spectrum of $-\Delta+V$ in $L^2(\R^n)$ is discrete then for every
function $\gamma:(0,+\infty)\rightarrow(0,1)$ and every $d>0$
\begin{equation}\label{inf_int}
\inf_{F\in\Nclass}\int_{\Gd\setminus
F}V(x)\,dx\rightarrow+\infty\quad\text{as}\quad
\Gd\rightarrow\infty.
\end{equation}

Sufficiency: Let a function $d\mapsto\gamma(d)\in(0,1)$ be defined for $d>0$ in a neighborhood of $0$ and satisfying
\[ \limsup_{d\downarrow0}d^{-2}\gamma(d)=+\infty. \]

Assume that there exists $d_0>0$ such that
\textrm{(\ref{inf_int})} holds for every $d\in(0,d_0)$. Then the
spectrum of $-\Delta+V$ in $L^2(\R^n)$ is discrete.
\end{teo}

\begin{rem}
It follows from the previous theorem that a necessary condition
for the discreteness of spectrum of $-\Delta+V$ is
\begin{equation}\label{necessary}
\int_{\Gd}V(x)\,dx\rightarrow\infty
\quad\text{as}\quad\Gd\rightarrow\infty.
\end{equation}
\end{rem}

The following lemma is very useful tool in  the next sections
\cite{GarciadelMoral:2006tb}.
\begin{lema}\label{lema}
For each given $\Gd=\Gd(x_0)$,
\[ \cd:=\inf_{F\in\Nclass}\vol{\Gd\setminus F}>0. \]
\end{lema}
\begin{proof}
Let $V(x)=|x|$. Then by Friedrichs theorem the spectrum of
$-\Delta+V$ is discrete, so by theorem \ref{teo_mazya} we have
\[ \inf_{F\in\Nclass}\int_{\Gd\setminus F}V(x)\,dx\to\infty\quad\text{as}\quad |x_0|\to\infty. \]
Now $\int_{\Gd\setminus F}V(x)\,dx\leq(|x_0|+d)\vol{\Gd\setminus
F}$ implies that
\[ \inf_{F\in\Nclass}\int_{\Gd\setminus F}V(x)\,dx \leq (|x_0|+d)\inf_{F\in\Nclass}\vol{\Gd\setminus F}, \]
from which follows that $\inf_{F\in\Nclass}\vol{\Gd\setminus
F}>0$, as we claimed.
\end{proof}
The following proposition extends the result of B. Simon in
\cite{Simon:1983B} which is used also as a toy model for the
membrane in \cite{deWit:1988ct}.
\begin{prop}
Let $V(x)=\dis\prod_{k=1}^n|x_k|^{\alpha_k}$, where $\alpha_k>0$
for all $k=1,2,\ldots,n$. Then the spectrum of the \schr\ operator
$-\Delta+V$ in $L^2(\R^n)$ is discrete.
\end{prop}
See\cite{GarciadelMoral:2006tb} for a proof using Molchanov ideas.
%%%%%%%%%%%%%%%%%%%%%%%%%%%%%%%%%%%%%%%%%%%%%%%%%%%%%%%%%%%%%%%%%%%%%%%%%%%%%%%%%%%%%%%%%%%%%%%%%%%%%%%%%%%%%%%%%%%%%%%%%%%%%%%%%%%%%%%

\section{Uniformly Bounded Basis}

In this section we prove that the orthonormal basis for polynomials
with respect to the inner product over $\Omega_F$ is uniformly bounded,
independently of $F$ in $\mathcal{G}_D$. This result will be used in
the main proof of the paper.
\begin{prop}
Let $\R[x_1,\ldots, x_M]$ be the ring of polynomials over $\R$ in $M$ indeterminate and let  $\mathcal{P}=\mathrm{span}\{%\varphi
P_k(x)\in\R[x_1,\ldots, x_M],\ k=1,\ldots,N\}$ be a subspace of dimension $N$. Let $F\in\Nclass$ and $\OF=\Gd\setminus F$, then
\[\|%\varphi
P_k\|_{\OF}^2:=\int_{\OF}%\varphi
P_k^2(x)\,dx>0,\quad\text{for all $k$}.\]
\end{prop}

\begin{proof}
There exists $c_d>0$ such that $\vol{\Gd\setminus F}\ge c_d$,
hence %there exists a non empty open ball  $B_F\subset\OF$ depending on $F$.
for each $F\in\Nclass$, there exists a non empty open ball $B_F\subset\OF$

Suppose that $\|%\varphi
P_k\|_{\OF}^2=0$. Then $P_k|_B(x)\equiv0$ (here $P_k|_B$ stands for the restriction of $P_k$
to the set $B$) for any open ball $B\subset\OF$. In particular, $P_k|_{B_F}(x)\equiv0$.

As $P_k$ is not the zero polynomial, we have $P_k(x)=a_\alpha x^\alpha+(\textrm{lower order terms})$ with $a_\alpha\ne0$ and $|\alpha|\ge0$.
%Let $B\subset\OF$ be a non empty open ball, then $\varphi_k|_B(x)=0$  and hence
Thus, $\dis\frac{\partial^{|\alpha|}}{\partial x^\alpha}P_k|_{B_F}(x)=a_\alpha\alpha!\ne0$, which
contradicts the fact that $P_k|_{B_F}(x)\equiv0$. Therefore $\|P_k\|_{\OF}^2>0$.
\end{proof}

\begin{rem}
The argument in the proof is essentially that on a bounded set, the Lebesgue measure of the zero set of a non zero polynomial is zero.
\end{rem}

Let $\mathcal{B}=\{P_k(x)\}_{k=1}^N$ be a basis of $\mathcal{P}$ (we are keeping the previous notations).
Following the Gram-Schmidt process, with respect to the inner product
\[ (f,g)_F:=\int_{\OF} f(x)g(x) dx\quad(f,g\in\mathrm{C}(\Gd)), \]
we can get an orthonormal basis $\{\varphi_m^F(x)\}_{m=1}^N$ for the space $\mathcal{P}$.
It is clear that we can write $\varphi_m^F(x)=\sum_{k=1}^mb_{mk}^F P_k(x)$, i.e., $b_{mk}^F=0$ if $k>m$.

Now we can  state the main result of this section:
\begin{teo}\label{main-thm}
There exists $C>0$ independent of $F$ such that $|\varphi_m^F(x)|\le C$ for all $x\in\OF$ and all $F\in\Nclass$.
\end{teo}

In order to prove this theorem we need to establish the following preliminaries results.

In the previous notation, let $b_m^F=(b_{m1}^F,\ldots,b_{mm}^F,0,\ldots,0)^\top\in\R^N$. Then
\begin{enumerate}
\item $\{b_m^F\}_{k=1}^N$ is a basis of $\R^N$ and the application $P_m^F\mapsto b_m^F$ defines an isomorphism between $\mathcal{P}$ and $\R^N$.
\item  Moreover, if $\Phi^F\in\R^{N\times N}$ is the matrix given by $\Phi^F_{kj}=\dis\int_{\OF}P_k(x)P_j(x)\,dx$, then
\begin{enumerate}
\item[(2.1)] $\delta_{mn}=(\varphi^F_m,\varphi^F_n)_F=\langle \Phi^Fb_m^F,b_n^F\rangle:=(\Phi^Fb_m^F)^\top b_n^F$ (the Kronecker delta).
\item[(2.2)] $\Phi^F$ is symmetric and positive definite. Hence the eigenvalues of $\Phi^F$ are positive.
\end{enumerate}
\end{enumerate}

\begin{proof}
These assertions follow from straightforward arguments. Let $u\in\R^N$ then $u=\sum_{k=1}^Na_mb_m^F$. The last statement of (2.2) follows from
\[\langle\Phi^Fu,u\rangle=\sum_{m,n=1}^Na_ma_n\langle \Phi^Fb_m^F,b_n^F\rangle=\sum_{m=1}^Na_m^2\ge0,\]
with equality only in the trivial case.
\end{proof}

\begin{lema}
Let $\sigma(\Phi^F)$ be the spectrum of $\Phi^F$, and let $\lambda_F:=\min\sigma(\Phi^F)$. Then \[\inf_{F\in\Nclass}\lambda_F>0.\]
\end{lema}
\begin{proof}
Suppose that $\dis \inf_{F\in\Nclass}\lambda_F=0$. Then there is a sequence of matrices $\Phi^{F_1},\Phi^{F_2},\ldots$ with $F_n\in\Nclass$
and there are sequences of eigenvalues and eigenvectors  $\lambda_{F_1},\lambda_{F_2},\ldots$ and $u_1,u_2,\ldots$, respectively, with $\Phi^{F_n}u_n=\lambda_{F_n}u_n$, and such that
\[\lambda_{F_n}\to0\quad\textrm{ as }\quad n\to\infty.\]

{\bf I.} Without loss of generality, we can suppose that $\|u_n\|=1$ for all $n$. Then, using that the unit sphere
is compact in  finite dimensions, there exists a convergent subsequence with a limit unit vector $u_0$. In order to
simplify notation, suppose that $u_n\to u_0$. Let $u_n=(u_{n1},\ldots,u_{nN})^\top,\ n=0,1,2,\ldots$, and let $\psi_n=\sum_{k=1}^Nu_{nk}P_k$. Then
\[ \|\psi_n\|^2_{\OFn}=\int_{\OFn}\left(\sum_{k=1}^Nu_{nk}P_k\right)^2\,dx=\langle \Phi^{F_n}u_n,u_n\rangle=\lambda_{F_n}\to 0 \textrm{ as } n\to\infty. \]

{\bf II.} Let $\vec{P}(x)=(P_1(x),\ldots,P_N(x))^\top$, and let
$\psi(u,x):=\langle
u,\vec{P}(x)\rangle=u\cdot\vec{P}(x)=\sum_{k=1}^Nw_kP_k(x)$ (where
$u=(w_1,w_2,\ldots,w_N)$). Then $\psi(u,x)$ is a polynomial and
$\psi(u,x)\in\mathcal{P}$. Therefore the zero set of $\psi(u,x)$,
namely $Z(\psi(u,x))=\{x\in\R^N: \psi(u,x)=0\}$ has measure zero for
all $u\in\R^N$. Let $\mathsf{N}_0$ be an open neighborhood of
$Z(\psi(u_0,x))\cap\Gd$ with measure $c_d/2$ (or more). \vskip 0.5cm

$\psi(u_0,x)^2$ is a continuous function, hence on the compact set
$\overline{\mathcal{G}_D\setminus\mathsf{N}_0}$ it has a minimum
$m^2\ne 0$. We denote $M^2=\int_{\mathcal{G}_D}\|\vec{P}(x)\|^2>0.$ We
have, recalling that $\psi_n=\psi(u_n,x)$,
\begin{align*}
\|\psi_n\|^2|_{\OFn}\ge & 
\|\psi_n\|^2_{\Omega_{F_n}\setminus\mathsf{N}_0}=\|\psi(u_0,x)+\psi(u_{n}¥-u_0,x)\|^2_{\Omega_{F_n}\setminus\mathsf{N}_0}\ge\\
& \vert\quad 
\|\psi(u_0,x))\|_{\Omega_{F_n}\setminus\mathsf{N}_0}-\|\psi(u_{n}¥-u_0,x)\|_{\Omega_{F_n}\setminus\mathsf{N}_0}\vert^2.
\end{align*}
and
\begin{align*}
 \|\psi(u_{n}¥-u_0,x)\|_{\Omega_{F_n}\setminus\mathsf{N}_0}^2\le \| u_{n}¥-u_0\|^2 M^2.
\end{align*}
Consequently,
\begin{align*}
 \|\psi_n\|^2_{\Omega_{F_n}\setminus\mathsf{N}_0}\ge \frac{1}{8}m^2 c_d
\end{align*}
for all $u_{n}¥$ such that $\|u_{n}¥-u_0\|^2\le 
\frac{1}{8}\frac{m^2}{M^2}c_d$, which is a contradiction with conclusion of (I),
which was a consequence of the assumption $\inf_{F\in\mathcal{N}_\gamma (\mathcal{G}_0,R^M)}\lambda_F=0$.

%%%%%%%%%%%%%%%%%%%%%%%%%%%%%%%%%%%%%%%%%%%%%%%%%%%%%%%%%%%%%%%%%%%%%%%%%%%%%%%%%%%%%%%%%%%%%%%%%%%%%%%%%%%%%%%%%%%%%%%%
Therefore
\[\inf_{F\in\Nclass}\lambda_F>0.\]
\end{proof}

\begin{corollary}
There exists $K>0$ independent of $F\in\Nclass$ such that $\|b_m^F\|\le K$ for all $m=1,\ldots, N$.
\end{corollary}
\begin{proof}
Let $K_1=\inf_{F\in\Nclass}\lambda_F$. As $\Phi^F$ is symmetric,
there exists $S_F\in\R^{N\times N}$ orthogonal such that
$\Phi^F=S_F^\top D_FS_F$ where
$D_F=\textrm{diag}(\lambda_1,\ldots,\lambda_N)$ with
$\{\lambda_k\}_{k=1}^N=\sigma(\Phi^F)$.

Fix $m$ and let $w=S_Fb_m^F=(w_j)^\top$. Then $\|w\|=\|b_m^F\|$ and
\[1=(\varphi^F_m,\varphi^F_m)_F=\langle\Phi^Fb_m^F,b_m^F\rangle=(S_Fb_m^F)^\top D_FS_Fb_m^F=\sum_{j=1}^N\lambda_jw_j^2\ge
\lambda_F\|b_m^F\|^2\ge K_1\|b_m^F\|^2.\]
Hence $\dis \|b_m^F\|\le\frac{1}{\sqrt{K_1}}:=K.$
\end{proof}

Now we can prove the main theorem.
\begin{proof}{\bf (Theorem \ref{main-thm})}
There exists $C_0>0$ such that $|P_k(x)|\le C_0$ for all $k=1,\ldots, N$ and all $x\in\Gd$.
Let $x\in\OF$ and let $K$ be the constant in the previous corollary. Then
\[|\varphi_m^F(x)|=\left|\sum_{k=1}^mb_{mk}^FP_k(x)\right| \le \sum_{k=1}^m|b_{mk}^F||P_k(x)|\le m\|b_m^F\|C_0\le NKC_0:=C.\]
\end{proof}

\section{ Discreteness of the spectrum of some Schr\"{o}edinger 
operators with polynomial potentials.}

In this section we prove two propositions ensuring the discreteness of the
spectrum of some Schr\"{o}edinger operators with positive polynomial
potentials.

\begin{prop}\label{pro4.1}
Let%
\begin{equation}
V(X)=\sum \limits_{j=1}^{J}P_{j}^{2}(X)
\end{equation}%
Where $P_{j}$ belong to the ring of polynomials over $R$ in $M$ variables
and span a nontrivial subspace of it. If the Schrodinger operator $-\Delta
+V(X)$ has discrete spectrum in $L^{2}(R^{M})$, then the operator $-\Delta +%
\sqrt{V(X)}$ has also discrete spectrum in $L^{2}(R^{M}).$
\end{prop}

\begin{proof}
Let $\mathcal{G}_{d}=\mathcal{G}_{d}(X_{0})\subset \mathbb{R}^{M\text{ }}$be a ball centered at $X_{0}$ and radius $d>0$, let $F\in
\mathcal{N}_{\gamma }(\mathcal{G}_{d};\mathbb{R}^{n})$. We decompose $X=X_{0}+\xi $ for all $X$ in the cell $\mathcal{G}%
_{d}\backslash F$. Let $\Omega _{F}$ be the set of all such $\xi $.
Then the necessary condition of Theorem 1 implies that

\begin{equation}
inf_{F\in
\mathcal{N}_{\gamma }(\mathcal{G}_{d};\mathbb{R}^{n})}\int_{\Omega _{F}}V(X_{0},\xi )d\xi \longrightarrow \infty \text{ as }%
\left\vert X_{0}\right\vert \longrightarrow \infty   \label{E1}
\end{equation}

We can rewrite the potential as%
\begin{equation}
V(X)=\sum\limits_{j=1}^{J}P_{j}^{2}(X_{0},\xi )  \label{E2}
\end{equation}%
where $P_{j}(X_{0},\xi )$, are polynomials in $\xi $ with
coefficients depending on $X_{0}$. Let us denote by $N$ the
dimension of the subspace span by $P_{j}(X_{0},\xi )$ with
$j=1,...,J$. From this set we consider $N$ independent polynomials,
and following the Gram-Schmidt process, whit
respect to the inner product%
\begin{equation}
(f,g)_{F}=\int_{\Omega _{F}}f(\xi )g(\xi )d\xi   \label{E3}
\end{equation}%
we can get an orthonormal basis $\varphi _{k}^{F}\left( \xi \right) $ 
with $%
k=1,...,N$ for the subspace span by $P_{j}(X_{0},\xi )$, $j=1,...,J$. It is
possible to write

\begin{equation*}
P_{j}(X_{0},\xi )=\sum\limits_{k=1}^{N}a_{jk}(X_{0})\varphi _{k}^{F}(\xi )%
\text{ \ \ \ \ \ }j=1,...,J.
\end{equation*}

where $a_{jk}(X_{0})$ depends on the set $\Omega _{F}.$

Let us denote by $M_{F}>0$ a uniform bound for $\varphi _{k}^{F}\left( \xi
\right) $ such that $\left\vert \varphi _{k}^{F}\left( \xi \right)
\right\vert \leq M_{F}$ for all $k$ and all $\xi \in \Omega _{F}.$

We have,%
\begin{equation}
\left\Vert P_{j}\right\Vert _{\Omega _{F}}^{2}:=\int_{\Omega
_{F}}P_{j}^{2}(X_{0},\xi )=\sum\limits_{k=1}^{N}a_{jk}^{2}(X_{0})\text{ and
}\int_{\Omega _{F}}Vd\xi =\sum_{j}\left\Vert P_{j}\right\Vert _{\Omega
_{F}}^{2}  \label{E4}
\end{equation}%
Then using $\left( \sum\limits_{k=1}^{n}a_{k}\right) ^{2}\leq
n\sum\limits_{k=1}^{n}a_{k}^{2}$ twice, we have $P_{j}^{2}\leq
N^{3}\sum\limits_{k=1}^{N}a_{jk}^{4}\varphi _{k}^{4^{F}}(\xi )$, therefore%
\begin{equation*}
\int_{\Omega _{F}}P_{j}^{4}d\xi \leq
N^{3}\sum\limits_{k=1}^{N}a_{jk}^{4}\int_{\Omega _{F}}\varphi
_{k}^{4^{F}}(\xi )d\xi \leq
N^{3}M_{F}^{2}\sum\limits_{k=1}^{N}a_{jk}^{4}\leq N^{3}M_{F}^{2}\left(
\sum\limits_{k=1}^{N}a_{jk}^{2}\right) ^{2}
\end{equation*}%
i.e. $\int_{\Omega _{F}}P_{j}^{4}d\xi \leq N^{3}M_{F}^{2}\left\Vert
P_{j}\right\Vert _{\Omega _{F}}^{4}.$ Then from this, $(\ref{E2})$ and $(\ref%
{E4})$%
\begin{equation}
\int_{\Omega _{F}}V^{2}(X_{0},\xi )d\xi \leq
N^{4}M_{F}^{2}\sum\limits_{j}\left\Vert P_{j}\right\Vert _{\Omega
_{F}}^{4}\leq N^{4}M_{F}^{2}\left( \int_{\Omega _{F}}V(X_{0},\xi )d\xi
\right) ^{2}  \label{E5}
\end{equation}%
Since $V^{\alpha }\in L^{2}\left( \Omega _{F}\right) $ for all real $\alpha
\geq 0$, using Schwarz inequality twice, we obtain:%
\begin{equation}
\left( \int_{\Omega _{F}}Vd\xi \right) ^{\frac{3}{2}}\leq \int_{\Omega
_{F}}V^{\frac{1}{2}}d\xi \left( \int_{\Omega _{F}}V^{2}d\xi \right) ^{\frac{1%
}{2}}  \label{E6}
\end{equation}%
Now using $(\ref{E5})$ and $(\ref{E6})$ we have:

\begin{equation*}
\left( \int_{\Omega _{F}}Vd\xi \right) ^{\frac{1}{2}}\leq
N^{2}M_{F}\int_{\Omega _{F}}V^{\frac{1}{2}}d\xi
\end{equation*}%
and from Theorem 7 $M_{F}\leq \mathbb{C}$ independent of $F$.
Consequently using ($\ref{E1}$) and the sufficient condition of
Theorem 1, we conclude that $-\Delta +\sqrt{V(X)}$ has discrete
spectrum.
\end{proof}

\begin{corollary}
Let $V(X)$ be as in Proposition \ref{pro4.1}. If $-\Delta +V(X)$ has discrete spectrum
in $L^{2}(R^{M})$, then $-\Delta +V(X)^{\frac{1}{2n}}$ for $n\geq 1$ natural
number, has also discrete spectrum in $L^{2}(R^{M})$.
\end{corollary}
%%%%%%%%%%%%%%%%%%%%%%
\begin{proof}
From the two previous inequalities we obtain

\begin{equation*}
\left( \int_{\Omega _{F}}V^{\frac{1}{2n}}\right) ^{\frac{1}{2}}\leq\mathbb{C}_{n}\int_{\Omega _{F}}V^{\frac{1}{4n}}
\end{equation*}%
for some $\mathbb{C}_{n}>0.$ It implies the above result for all $n\geq 1$.
\end{proof}

In the next proposition we use the following notation:

\begin{equation}
\sum\limits_{M_{1},...,M_{l}}:=\sum\limits_{M_{1}=1}^{M}\sum%
\limits_{M_{2}=1}^{M}...\sum\limits_{M_{l}=1}^{M}  \label{E7}
\end{equation}

\begin{equation}
{\sum\limits_{M_{1},...,M_{l}}}^{\prime}:=\sum\limits_{M_{1}}^{M}\sum%
\limits_{M_{2}}^{M}...\sum\limits_{M_{l}}^{M} \quad
\textrm{with}\quad  M_{1}\neq M_{2}\neq ...\neq M_{l}. \label{E8}
\end{equation}%

Given an index $M_{l}=k\in 1,2,\ldots M$ then

\begin{equation}
{\sum\limits_{M_{1},...,M_{l-1}}}^{\prime\prime}:=\sum\limits_{M_{1}}^{M}\sum%
\limits_{M_{2}}^{M}...\sum\limits_{M_{l-1}}^{M} \quad
\textrm{with}\quad  M_{1}\neq M_{2}\neq ...\neq M_{l-1} \neq k \label{E9}
\end{equation}%

Given a set of real coefficients $f_{a_{1},...,a_{l}}^{B}$ where $%
a_{1},...,a_{l}=1,...,N$ we denote

\begin{eqnarray}\label{E10}
&&\mathcal{F}_{a_{1},...,a_{l-1};\widehat{a}_{1},...,\widehat{a}_{l-1}}
=f_{ca_{1},...,a_{l-1}}^{B}f_{c\widehat{a}_{1},...,a_{l-1}}^{B}+
\\ \nonumber &&....+f_{a_{1},...,a_{i-1},c,a_{i+1},..,a_{l-1}}^{B}f_{\widehat{a_{1}},..,
\widehat{a}_{i-1},c,\widehat{a}_{i+1},..,\widehat{a}_{l-1}}^{B}+...
...+f_{a_{1}...a_{l-1}c}^{B}f_{\widehat{a_{1}},...,\widehat{a}%
_{i-1},c,}^{B}
\end{eqnarray}
 and in general
\begin{equation}
\mathcal{F}_{a_{1},...,a_{l-i};\widehat{a}_{1},...,\widehat{a}_{l-i}}:=\sum
f_{(c_{1},....,c_{i};a_{1},...,a_{l-i})}^{B}f_{(c_{1},...,c_{i};\widehat{a}%
_{1},...,\widehat{a}_{l-i})}^{B}  \label{E11}
\end{equation}
where $(c_{1},...,c_{i};a_{1},...,a_{l-i})$ denotes a set of $l$ indices. $i$
indices are $c_{1},...,c_{i}$ in that order and $l-i$ indices are $%
a_{1},...,a_{l-i}$ in that order, the summation is in all possible
sets. $B$ is an independent set of indices  with any range over which a sum is
performed.

We also introduce the matrix $\mathcal{M}$ with components

\begin{equation}
M_{a\widehat{a}}:=\mathcal{F}_{a;\widehat{a}%
}=f_{c_{1},...,c_{l-1},a}^{B}f_{c_{1},...,c_{l-1},\widehat{a}%
}^{B}+....+f_{c_{1},...,c_{i-1},a,c_{i+1},...c_{l-1}}^{B}f_{c_{1},...,c_{i-1},%
\widehat{a},c_{i+1},...,c_{l-1}}^{B}+...  \label{E12}
\end{equation}

In the proof of the next proposition we will use the following
remark.

%\begin{prop}\label{pro4.2}
\begin{rem}
If $H=-\Delta +X^{c}X^{\hat{c}}L_{c\widehat{c}}$, where $\left[ L_{c\widehat{%
c}}\right] $ is symmetric and positive then $H\geq \sqrt{tr\left[ L_{c%
\widehat{c}}\right] }.$ In fact $\left[ L_{c\widehat{c}}\right] =S^{T}DS$
may be diagonalized and $-\Delta $ is invariant under a rotation $%
X\rightarrow Y=SY$. \ We then have\qquad \qquad \qquad \qquad \qquad \qquad
\qquad \qquad \qquad \qquad \qquad \qquad \qquad \qquad \qquad \qquad \qquad
\qquad \qquad \qquad \qquad \qquad \qquad \qquad \qquad \qquad \qquad \qquad
\qquad \qquad \qquad \qquad \qquad \qquad \qquad \qquad \qquad \qquad \qquad
\qquad \qquad \qquad \qquad \qquad \qquad \qquad \qquad \qquad \qquad \qquad
\qquad \qquad \qquad \qquad \qquad \qquad \qquad \qquad \qquad \qquad \qquad
\qquad \qquad \qquad \qquad \qquad \qquad \qquad \qquad \qquad \qquad \qquad
\qquad \qquad \qquad \qquad \qquad \qquad \qquad \qquad \qquad \qquad \qquad
\qquad \qquad \qquad \qquad \qquad \qquad \qquad \qquad \qquad \qquad \qquad
\qquad \qquad \qquad \qquad \qquad \qquad \qquad \qquad \qquad \qquad \qquad
\qquad \qquad \qquad \qquad \qquad \qquad \qquad \qquad \qquad \qquad \qquad
\qquad \qquad \qquad \qquad \qquad \qquad \qquad \qquad \qquad \qquad \qquad
\qquad \qquad \qquad \qquad \qquad \qquad \qquad \qquad \qquad \qquad \qquad
\qquad \qquad \qquad \qquad \qquad \qquad \qquad \qquad \qquad \qquad \qquad
\qquad \qquad \qquad \qquad \qquad \qquad \qquad \qquad \qquad \qquad \qquad
\qquad \qquad \qquad \qquad \qquad \qquad \qquad \qquad \qquad
\begin{equation}
H=\sum\limits_{m}-\frac{\partial ^{2}}{\partial y^{m^2}}+\lambda
_{m}y_{m}^{2}\geq \sum\limits_{i}\sqrt{\lambda _{i}}\geq \sqrt{tr\left[ L_{c%
\widehat{c}}\right] }
\end{equation}
\end{rem}
%\end{prop}

\begin{prop}\label{a1}
Let $H=-\Delta +V(X)$ be a Schr\"{o}edinger operator with potential $V(X)$
given by%
\begin{equation}
V(X)=\sum\limits_{M_{1},...,M_{l}}\sum\limits_{B=1}^{N}\left(
X_{M_{1}}^{a_{1}}...X_{M_{l}}^{a_{l}}f_{a_{1}...a_{l}}^{B}\right)
^{2} \label{E14}
\end{equation}%
let $\mathcal{M}$ be the symmetric matrix defined in (\ref{E12}),
$\left[ X_{M_{i}}^{a_{i}}\right] \in \mathbb{R}^{M\times N}$ and
$f_{a_{1},...,a_{l}}^{B}$ real coefficients satisfying the following
restriction: $\mathcal{M}$ is strictly positive definite. Then $H$
is essentially self adjoint and has a discrete spectrum in
$L^{2}\left( \mathbb{R}^{M\times N}\right) $.
\end{prop}

\begin{rem}
There is no assumption concerning the symmetry or antisymmetry of
$f_{a_{1},...,a_{l}}^{B}$ on the indices $a_{1},...,a_{l}$. It will
be clear from the following proof that instead of one $B$ index we
may have any number of them.
\end{rem}

\begin{proof}
We obtain the following inequalities

\begin{equation*}
V\left( X\right) \geq
{\sum\limits_{M_{1},...,M_{l}}}^{\prime} \sum\limits_{B}\left(
X_{M_{1}}^{a_{1}}...X_{M_{l}}^{a_{l}}f_{a_{1}...a_{l}}^{B}\right)
^{2}\geq
\mathit{k }_{0}\sum\limits_{M_{l}=1}^{M}X_{M_{l}}^{c}X_{M_{l}}^{\hat{c}%
}G_{c\widehat{c}}^{M_{l}}
\end{equation*}

for some real number $\mathit{k }_{0}>0$, where
\begin{equation*}
G_{c\widehat{c}}^{M_{l}}={\sum
\limits_{M_{1},...,M_{l-1}}}^{\prime\prime}X_{M_{1}}^{a_{1}}...X_{M_{l-1}}^{a_{l-1}}%
\mathcal{F}_{a_{1},...,a_{l-1};\widehat{a}_{1},...,\widehat{a}%
_{l-1}}X_{M_{1}}^{\hat{a}_{1}}...X_{M_{l-1}}^{\hat{a}_{l-1}}
\end{equation*}
does not depend on $X_{M_{l}}^{c}$.

For each $M_{l}$ we have a quadratic potential, we may then use
(\ref{E14}) to obtain
\begin{equation}
-\Delta +V(X)\geq \lambda _{o}\left( -\Delta +\mathbb{C}_{o}\sqrt{V_{1}\left( X\right) }\right)   \label{E15}
\end{equation}%
for some real $\lambda _{0}>0$ and $\mathbb{C}_{o}>0$. \ In the same way
\begin{equation}
-\Delta +V_{1}(X)\geq \lambda _{1}\left( -\Delta +\mathbb{C}_{1}\sqrt{V_{2}\left( X\right) }\right)   \label{E16}
\end{equation}

where%
\begin{equation}
V_{1}\left( X\right)
=\sum\limits_{M_{1},...,M_{l-1}}X_{M_{1}}^{a_{1}}...X_{M_{l-1}}^{a_{l-1}}%
\mathcal{F}_{a_{1},...,a_{l-1};\widehat{a}_{1},...,\widehat{a}%
_{l-1}}X_{M_{1}}^{\hat{a}_{1}}...X_{M_{l-1}}^{\hat{a}_{l-1}}  \label{E17}
\end{equation}

\begin{equation}
V_{2}\left( X\right)
=\sum\limits_{M_{1},...,M_{l-2}}X_{M_{1}}^{a_{1}}...X_{M_{l-2}}^{a_{l-2}}%
\mathcal{F}_{a_{1},...,a_{l-2};\widehat{a}_{1},...,\widehat{a}%
_{l-2}}X_{M_{1}}^{\hat{a}_{1}}...X_{M_{l-2}}^{\hat{a}_{l-2}}  \label{E18}
\end{equation}%
and in general

\begin{equation}
-\Delta +V_{i}(X)\geq \lambda _{i}\left( -\Delta +\mathbb{C}_{i}\sqrt{V_{i+1}}\right)   \label{E19}
\end{equation}

\begin{equation*}
V_{i}\left( X\right)
=\sum\limits_{M_{1},...,M_{l-i}}X_{M_{1}}^{a_{1}}...X_{M_{l-i}}^{a_{l-i}}%
\mathcal{F}_{a_{1},...,a_{l-i};\widehat{a}_{1},...,\widehat{a}%
_{l-i}}X_{M_{1}}^{\hat{a}_{1}}...X_{M_{l-i}}^{\hat{a}_{l-i}}
\end{equation*}
for some real numbers $\lambda _{i}>0$, $\mathbb{C}_{i}>0$, $i=0,...,l-2$.

For $i=l-2$ we get
\begin{equation}
-\Delta +V_{l-2}(X)\geq \lambda _{l-2}\left( -\Delta +\mathbb{C}_{l-2}\sqrt{V_{l-1}}\right)   \label{E20}
\end{equation}%
and
\begin{equation*}
V_{l-1}\left( X\right) =\sum\limits_{M_{1}}X_{M_{1}}^{a_{1}}\mathcal{F}%
_{a_{1};\widehat{a}_{1}}X_{M_{1}}^{\hat{a}_{1}}=\sum%
\limits_{M_{1}}X_{M_{1}}^{a}M_{a\widehat{a}}X_{M_{1}}^{\hat{a}}
\end{equation*}
\end{proof}

This is the potential of an harmonic oscillator in
$\mathbb{R}^{M\times N}$, since under the assumption of the
Proposition \ref{a1} $\mathcal{M}$ is strictly positive.
Consequently
\begin{equation*}
-\Delta +V_{l-1}(X)
\end{equation*}
has a discrete spectrum in $L^{2}\left( \mathbb{R}^{M\times
N}\right) .$ Using now Proposition (\ref{pro4.1}) we obtain that
$-\Delta +\mathbb{C}_{l-2}\sqrt{V_{l-1}}$ has discrete spectrum in
$L^{2}\left( \mathbb{R}^{M\times N}\right) $ and from (\ref{E20})
$-\Delta +V_{l-2}(X)$ has also discrete spectrum in
$L^{2}\left(\mathbb{R}^{M\times N}\right)$.

Using this argument several times we conclude that
\begin{equation*}
-\Delta +V(X)
\end{equation*}
has a discrete spectrum in $L^{2}\left( \mathbb{R}^{M\times N}\right) $.

The property of being essentially self adjoint arises from general
arguments for symmetric operators. Moreover $H$ is a positive
operator densely defined in $L^2(\mathbb{R}^{M\times N})$, then
there exists a positive self adjoint extension of $H$. It is called
the Friedrichs extension of $H$.

\section{Connection with ABJM-like theories}

In the previous section we rigorously showed  at non-perturbative
level, a sufficient condition for the discreteness of the spectrum of a Schr\"oedinger operator with a scalar polynomial potential of
any degree that can be expressed as the sum of squares. This result
generalizes all previously ones in the literature. The requirements
for the discreteness are very general and not restricted only to
cases of Lie groups or Fillipov algebras
 expressible as a direct product of Lie algebras, as discussed
below. The $F's$ are any kind of constant that satisfy the
regularity condition  stated in proposition (\ref{a1}).

This result, which holds for a class of scalar potentials, is far
from obvious. There is a fairly widespread belief that positive
definite polynomial scalar potentials, which can be expressed as a sum of squared terms, automatically have discrete spectrum. But this is not true.
%%%%%%%%%%%%%%%%%%%%%%%%%%%
We presented in the introduction an infinite class of such
potentials. Consider for instance, the well-known situation for the
$D=11$ supermembrane. The classical theory have unstable solutions
with minimal energy. The bosonic spectrum is discrete, in spite of
the fact that there are flat directions in the potential (the
spectrum is discrete because the Molchanov mean value goes to infinity
when one goes to infinity on the configuration space). The
supersymmetric spectrum is continuous
 from $(0,\infty)$ because the contributions from the spinor terms give rise to an unbounded from below potential,
 which although is balanced by the $-\Delta$ contribution (the hamiltonian is positive) it allows the wave function to escape to infinity. \newline

It is interesting that there are well-defined sectors of the supermembrane theory
(with topological central charge different from zero) which have a discrete spectrum from $(0,\infty)$
 with isolated eigenvalues with finite multiplicity \cite{Boulton:2002br}\cite{Boulton:2004yt},\cite{Boulton:2006mm}. We will show in this section that the scalar
 bosonic potentials for the M-branes, BLG, ABJM and ABJ theories, all have an associated Schr\"oedinger
 operator with discrete spectrum from zero to infinity, with isolated eigenvalues which have finite multiplicity.
 We will not consider in this paper the contribution to the potential arising from the Chern-Simons (CS) terms.
 The qualitative properties of the spectrum of BLG, ABJ/M supersymmetric models are so far unknown. There are three important
 properties in the general analysis of these last three theories. First the behaviour of the scalar bosonic potential
  (which is studied in this paper), second the CS contribution, and third the presence of sypersymmetry.
  The supersymmetric interacting terms containing spinor fields in these theories are quadratic on the fields
  in distinction with the case of the $D=11$ supermembrane in  the 
  Light Cone Gauge (L.C.G) where the depedence is linear.
  This point may have important consequences in a future analysis of the complete spectrum.\newline
\begin{itemize}
\item{\bf The BLG case.}
\end{itemize}

 To characterize  the non-perturbative spectral properties of
the scalar potential of BLG/ABJM type , it is necessary first, to
formulate these theories in the regularized matrix formalism. These
theories have real scalar fields  $X^{aI}$ valued in the
bifundamental representation of the $\mathcal{G}\times \mathcal{G'}$
algebra, gauge fields $A_{\mu}^{ab}$ where
    $\mu=0,1,2$ spanning the target-space dimensions, and $a\in 
    \mathcal{G},b\in \mathcal{G^{'}}$,
     and spinors $\Psi_{a\alpha }$ also valued in the algebra.
     Let us consider the sixth degree scalar potential of the BLG case,

    \begin{equation} V=\int dx^3
    \frac{1}{12}Tr([X^I,X^J,X^K])^2=\\
    \nonumber=\int dx^3\frac{1}{12}f^{abcd}f_d^{efg}(X^{I}_aX^{J}_bX^{K}_cX^{I}_eX^{J}_fX^{K}_g)\end{equation}
where $f_d^{efg}$ are the 'structure constants' of the algebra color 
generators $T_a$. For the BLG
case a 3-algebra relation is satisfied
\begin{equation} [T^a,T^b,T^c]=f^{abc}_{d}T^d.\end{equation}

We expand now each of the fields $X^{I}_a$ in a basis of generators
$T_A$ \footnote{In the case of the $D=11$ supermembrane a regularization procedure
in terms of $su(N)$ generators is natural because the structure constants of the $su(N)$
algebra converge in the large $N$ to the structure constants of the area preserving
 diffeomorphisms, the gauge symmetry of the $D=11$ Supermembrane in the light cone gauge. In the present
 case we do not have such argument however following other truncation procedures one can obtain the same result.
 A straightforward truncation procedure is to expand the fields on a orthonormal basis on the compact base manifold
 and truncate it at some level in order to have a model which can be analyzed by the rigorous methods of quantum mechanics.
 As we said the qualitative results are independent of these two regularization procedures.},
     to obtain the regularized model,
     \begin{equation} X^{I}_a= \sum X^{IA}_aT_A\end{equation} with
     $A=(a_1,a_2)$. For the enveloping algebra of $su(N$),
    \begin{equation}T_AT_B=h_{AB}^CT_C,\quad \eta_{AB}=\frac{1}{N^4}Tr{(T_AT_B)}\end{equation}  $h_{AB}^C$
    are given in \cite{deWit:1989vb},\cite{GarciadelMoral:2001zb}. We substitute the scalar potential by a
    regularized one
\begin{equation}
\begin{aligned}
V=&\frac{1}{12}f^{abcd}f^{efg}_d
 X^{IA}_aX^{BJ}_bX^{CK}_cX^{EI}_eX^{FJ}_fX^{GK}_gTr_{I_{N}\times I_{N}}(T_A
T_B T_C T_E T_F T_G)
\\ \nonumber &=\frac{1}{12}f^{abcd}f^{efg}_d h_{AB}^{U}h_{CE}^{V} h_{FG}^{W}h_{UVW}
 X^{IA}_aX^{BJ}_bX^{CK}_cX^{EI}_eX^{FJ}_fX^{GK}_g\end{aligned}\end{equation}
The potential can be re-written as a squared-term \begin{equation}
V=\frac{1}{12}(F^{\mathcal{A}\mathcal{B}\mathcal{C}}_{\mathcal{U}}X^{\mathcal{A}I}X^{\mathcal{B}J}X^{\mathcal{C}K})^2\end{equation}
with coefficients
$F^{\mathcal{A}\mathcal{B}\mathcal{C}}_{\mathcal{U}}=f^{abc}_{u}
h_{AB}^{E}h_{CE}^{U}$ that do not exhibit antisymmetry in the
indices $\mathcal{A}=(A,a),\mathcal{B}=(B,b),\mathcal{C}=(C,c)$ nor
are structure constants. Using the Proposition \ref{a1} we can
assure that this regularized potential has a purely discrete
spectrum since
\begin{equation}
f^{abcd}X_a^{IA}=0 \to X_a^{IA}=0.\Box
\end{equation}.
\newline
 The D=11 Supermembrane, the 5-brane, p-branes and
the N=8 Bagger Lambert model satisfy the regularity condition for
the matrix $\mathcal{M}$ of  Proposition \ref{a1}.
\begin{itemize}
\item{\bf The ABJ/M case}
\end{itemize}

ABJM theory can be obtained from the 3-algebra expression by
relaxing some antisymmetric properties of the 3-algebra structure
constant as it is indicated in \cite{Bagger:2008se} considering now
instead of real scalar fields, -as happens in the BLG case-, complex
ones $Z^{a\alpha }$ .\newline In the ABJM case
\cite{Aharony:2008ug}, the scalar potential may be re-expressed in a
covariant way as a sum of squares \cite{Bandres:2008ry}. Using the
results of \cite{Bagger:2008se} where the potential is
\begin{equation}
V = \frac{2}{3}\Upsilon^{CD}_{Bd}\bar\Upsilon_{CD}^{Bd} ,
\end{equation}
where
\begin{equation}
\Upsilon^{CD}_{Bd} =
f^{ab\overline{c}}{}_dZ^C_aZ^D_b\overline{Z}_{B\overline{c}}
-\frac{1}{2}\delta^C_Bf^{ab\overline{c}}{}_dZ^E_aZ^D_b\overline{Z}_{E\overline{c}}+\frac{1}{2}\delta^D_B
f^{ab\overline{c}}{}_dZ^E_aZ^C_b\overline{Z}_{E\overline{c}}.
\end{equation}
The zero-energy solutions correspond to $\Upsilon^{CD}_{Bd}=0$. In
distinction with the case of BLG, the ABJM potential includes a sum
of three squared terms. The indices $C,D$ are mandatory different
but not necessarily the index $B$. We can bound the potential for
the one with
\begin{equation}
\Upsilon^{CD}_{B^{'}d} = f^{ab\overline{c}}{}_dZ^C_a Z^D_b
\overline{Z}_{B^{'}\overline{c}}
\end{equation} where $B^{'}$ is an index different from $C,D$.
%%%%%%%%%%%%%%%%%%%%%%%%%%%%%%%%%%%%%%%%%%%%%%%%%%%

To reduce the analysis to one in quantum mechanics,  a regularization 
procedure is performed.  The regularity condition of Proposition
\ref{a1}, in terms of the triple product \cite{Bagger:2008se}, may
be expressed as a
\begin{equation}\label{a70}
[X,T^b;\overline{T}^{\overline{c}}]=f^{ab\overline{c}}{}_d X_a T^d
=0 \quad \forall{b,\overline{c}}\Rightarrow X_a=0.
\end{equation}
Note that if this condition is not satisfied, the
potentials we are considering have continuous spectrum. This result
follows using Molchanov, Maz'ya and Schubin theorem. Factorizing out the constants due to regularization process, in the case of
ABJM and ABJ it follows from (49) in \cite{Bagger:2008se} that
(\ref{a70}) implies\footnote{The condition holds in spite that the
polynomial expression considered in
 equation(49) of
\cite{Bagger:2008se} for the $f^{ab\overline{c}\overline{}d}$ in
terms of $u(N)$ generators is not the final one, since antisymmetry
in the two first indices still needs to be imposed by hand as the
authors explain. This fact does not affect the present condition
(\ref{ape}), since antisymmetric solutions represent a subset of
solutions we consider here.}
\begin{equation}\label{ape}
(t^{\lambda}_{\alpha})^{a \overline{c}}X_a=0,
\end{equation}
where $t^{\lambda}_{\alpha}$ are $u(N)$ representations of the gauge algebra $\mathcal{G}$.
 In the case $\mathcal{G}$ is $u(N)$ then the regularity condition is satisfied.

%%%%%%%%%%%%%%%%%%%%%%%%%%%%%%%%%%%%%%%%%%%%%%%%%%%%%%%%%%%%%%%%%%%%%%

The proposition (\ref{a1}) in our paper ensures then that the
Schr\"oedinger operator associated to the regularized scalar sixth
degree potential of ABJM has also purely discrete spectrum.$\Box$
\newline
 This  proves a  necessary condition for quantum
stability for the new supersymmetric models.  In fact, a continuous
spectrum at the regularized bosonic model arising from a formulation
on a compact space, would imply several difficulties on the models.
For example, the Feynmann kernel would be ill defined.  This is the
first step in order to consider a non perturbative analysis of these
new supersymmetric models.
\begin{itemize}
\item{\bf Some more comments.}
\end{itemize}
If we now add the regularized Chern-Simons gauge contribution  $
V_{CS}$ to the scalar potential there are quadratic and cubic
contributions. Without loose of generality, take for simplicity the
BLG case. Although the shape of CS terms clearly do not suit in the
shape of the potentials here considered, one could imagine to bound
this potential $V_{sixth}+V_{CS}$, however one can realize that the
cubic contribution is not necessarily positive. A gauge fixing
procedure must be performed here in order to analyze the problem. At
this stage we cannot guarantee the discreteness of the regularized
bosonic potential once the gauge fields are added and a further
study is needed in this approximation. However the main point
concerning the stability of the bosonic multiple branes is the
analysis of the $V_{sixth}$.
\newline
Another interesting issue is the spectral characterization of the
complete hamiltonians including their supersymmetric extension. The
analysis then, is much more involved. All of these actions of
multiple M2's have in common the construction of a conformal
supersymmetric gauge theory with quadratic couplings in the
fermionic variables. It goes like combinations of terms of the type
$ \bar{\Psi}^\dagger( \Gamma X X^{ \dagger}) \Psi$ . Their fermionic
contribution in the light cone Hamiltonian formulation
\cite{Nilsson:2008ri}, in distinction with the case of a single M2
brane, still depends quadratically on the bosonic variables. The
sufficient condition for discreteness of supersymmetric potentials
shown in \cite{Boulton:2002br} is no longer applicable and although
it does not exclude completely the possibility of the spectrum be
discrete at regularized non-perturbative level, makes it much more
fine tuned.
\section{Conclusions}
We obtain a general class of polynomials for which the Sch\"oedinger
operator has a discrete spectrum.
 This class includes all the scalar potentials in membrane, 5-brane, p-branes, multiple M2 branes,
 BLG and ABJM theories. We provide a proof of the discreteness of the spectrum of the associated
 Schr\"oedinger operators. This a a first step in order to analyse BLG and ABJM super-symmetric theories from a non-perturbative point of view.
 This  proves a  necessary condition for quantum
stability for the new supersymmetric models.  In fact, a continuous
spectrum at the regularized bosonic model would imply several
difficulties on the models. For example, the Feynmann kernel would
be ill defined.  This is the first step in order to consider a non
perturbative analysis of these new supersymmetric models.
%%%%%%%%%%%%%%%%%%%%%%%%%%%%%%%%%%%%

\section{Acknowledgements}
MPGM would like to thank I. Cavero-Pelaez, C.Y. Perez and M. Varela
for their help with the manuscript. The work of MPGM is funded by
the Spanish Ministerio de Ciencia e Innovaci\'on (FPA2006-09199) and
the Consolider-Ingenio 2010 Programme CPAN (CSD2007-00042). The work
of AR, IM, AP and LN are financed by  Decanato de Investigaciones y
Desarrollo(DID-USB), under Proyecto G-11 (AR and IM) and Proyecto 
S1-IC-CB-006-08  (AP and LN).

\bibliography{lista2}
\bibliographystyle{unsrt}
\end{document}